\documentclass[journal,10pt,twoside,twocolumn,letterpaper]{IEEEtran}
\usepackage{graphicx, epstopdf}
\usepackage{amssymb}
\usepackage{amsmath}
\usepackage{amsthm}
\usepackage{mathrsfs}
\usepackage{cite}
\usepackage{color}
\usepackage[acronym]{glossaries}
\usepackage[T1]{fontenc}
\usepackage[latin9]{inputenc}
\usepackage{array}
\usepackage{textcomp}
\usepackage{multirow}
\usepackage{soul}
\usepackage{flushend}
\makeatletter
\newcommand*\titleheader[1]{\gdef\@titleheader{#1}}
\AtBeginDocument{%
  \let\st@red@title\@title
  \def\@title{%
    \bgroup\normalfont\small\centering\@titleheader\par\egroup
    \vskip0.5em\st@red@title}
}
\makeatother
\titleheader{This is the authors'version of the paper that has been accepted for publication in IEEE Potentials Magazine}
\title{RF Wireless Power Transfer: Regreening Future Networks}
\author{
        {
        Ha-Vu Tran$^{1}$ and Georges Kaddoum$^{1}$
        }

\thanks{
$^1$Ha-Vu Tran and Georges Kaddoum are with University of
Qu\'{e}bec, \'{E}TS engineering school, LACIME Laboratory, 1100 Notre-Dame west, H3C 1K3, Montreal, Canada.
Email: \{ha-vu.tran.1@ens.etsmtl.ca, georges.kaddoum@etsmtl.ca.\}
This work has been supported by NSERC discovery grant 435243 - 2013.
}
 }

\begin{document}
    \maketitle

\begin{abstract}
Green radio communication is an emerging topic since the overall footprint of information and communication technology (ICT) services is predicted to triple between 2007 and 2020. 
Given this research line, energy harvesting (EH) and wireless power transfer (WPT) networks can be evaluated as promising approaches.
In this paper, an overview of recent trends for future green networks on the platforms of EH and WPT is provided.
By rethinking the application of radio frequency (RF)-WPT, a new concept, namely \textbf{\textit{green}} RF-WTP, is introduced.
Accordingly, opening challenges and promising combinations among current technologies, such as  small-cell, millimeter (mm)-wave, and Internet of Things (IoT) networks, are discussed in details to seek solutions for the question with how to re-green the future networks?

\end{abstract}


\section{Introduction}
Over past few years, green radio communication has drawn much attention from the research community, and it has strong impacts on various aspects, such as telecom business, wireless technologies, and natural environments. Specifically, the electricity cost and CO$_2$ emissions have been increasing due to wireless network operation.
For instance, the number of base stations (BSs) is more than 4 million, and each BS consumes an average of 25MWh per year (estimated approximate 80 percent of the total network's power consumption). Bearing in mind the environmental perspective, generating sufficient power to supply the networks causes a significant amount of CO$_2$ footprint. 
Particularly, the overall footprint of information and communication technology (ICT) services, e.g., computer, cell phone, and satellite networks, is predicted to triple by 2020 [1]. 

Recently, towards a future green world, energy harvesting (EH) technique has the potential to deal with the problem of energy inefficiency [2]-[5]. 
The main advantages of this approach can be presented in two-fold. First, the EH techniques harnesses green energy from natural sources, e.g., solar and wind. Thus, it contributes to reducing the overall footprint in order to protect surrounding environments. Nowadays, the popularity of using conventional energy sources, e.g., diesel, still dominates the use of the green sources. However, although the overall implementation cost of EH solutions is higher than that of conventional ones, this cost might be compensated after several years of operation. Second, another main challenge in future networks is prolonging the lifetime of smart user devices. Given this concern, EH networks take a tremendous advantage in various specific applications. For instance, EH is an efficient solution for reducing battery replacement costs in wireless sensor networks. Also, it can recharge the devices working in areas where the traditional power supply is infeasible, e.g., robotic devices working in toxic environments.
Nevertheless, the amount of harvested energy from natural resources, such as solar and wind, may vary randomly over time and depend on locations and weather conditions.
In other words, harvesting energy from these sources is not controllable and sustainable. For instance, there exists insufficient sunlight at night to generate energy, or it is difficult for indoor devices to harvest solar energy. In this context, radio frequency (RF) wireless power transfer (WPT) might be a promising approach to overcome such a drawback.

In this paper, we provide a comprehensive review to address the mentioned question of regreening the future world. 
In more specific details, the main contributions of this paper can be summarized as follows. 
First, by rethinking the use of RF-WPT, a concept so-called {\it green} RF-WTP is introduced. 
Second, a vision of future green networks based on the platforms of EH and green RF-WPT is presented. Thus, we discuss potential scenarios with the purpose of bridging green resources to indoor energy-hungry devices in the networks. 
Third, applications of various interesting concepts, such as small-cell, mm-wave, internet of things (IoT), etc., networks are highlighted. Given this concern, the challenges on each concept are identified.
Further, we discuss some attractive combinations of the existing concepts, such as a mixture of full-duplex, RF-WPT, and mm-wave, and investigate how the latter works together properly. To this end, promising trends in future are drawn to provide solutions for future re-greened networks.

\section{Energy Harvesting and Green RF Wireless Power Transfer}
In this section, an overview of EH models and a discuss on green RF wireless power transfer are shown.

\subsection{EH Models}
EH methodology might be described by harnessing energy from surrounding environments or thermal and mechanical sources, and then converting the latter into electrical energy. The generated electrical current can be used to supply devices by RF wireless power transfer. 
Generally speaking, EH models can be classified into two
architectures with harvest-use, and harvest-store-use.
In the first one, energy is harvested, and then is used instantly. Besides, given the second one, energy is harvested
as much as possible and then stored for future uses. In the following, we discuss the characteristics of such models.

In the harvest-use architecture, the EH systems directly supply devices. To guarantee the operation of the devices, the power output of the EH systems should be higher than the threshold of minimum working requirements. Otherwise, the devices would be disabled because there is not enough power supplied. As a consequence, unanticipated fluctuation in harvesting capacity close to threshold yields the working devices to vacillate in ON and OFF states.

Further, the harvest-store-use model includes a component storing harvested energy and also powering the connected devices.
Thanks to the storage, the energy can be harvested until sufficient for supplying the devices.
Moreover, such energy might be stored for later uses when there is lack of produced energy or the devices need to increase the working performance.
The storage component might include two parts of primary and secondary storages. In this context, the secondary storage can be seen as a backup one.
In particular, the harvest-store-use system can make non-stable but foreseeable energy sources, such as solar and wind, more favorable in uses.

\subsection{Green RF Wireless Power Transfer}
Over the last decade, solar, wind, mechanical, and thermal energy can be considered as the most efficient resources generating green energy usable for wireless networks.
However, the main drawback of such sources is the lack of stability.
In the quest for an alternative solution, the research community has explored that radio signals belonging to a frequency range from 300 GHz to 3 kHz can be used to carry energy over the air [3], [6], [7].
 On this basis, a transmitter can proactively recharge wireless devices by sending energy-bearing RF signals whenever it is necessary. This is the principle of a so-called RF-WPT technique.

In fact, it is well-known that EH is a green technique since it helps to reduce the footprint. However, in a shared vision, the RF-WPT technique seems to be harmful to surrounding environments because it costs electricity to generate RF signals, and causes electromagnetic pollution to the human body as well as interferences to data transmission. 
By rethinking the role of RF-WTP technique, 
we suggest that the RF-WTP can be seen as green {\it iff} (i) the RF signal carrying energy is generated using power harvested from green resources (for example, BSs are connected with outdoor energy harvesters to harvest-and-store green energy and then use such energy to wirelessly recharge to indoor devices using RF signals), and (ii) a tight restriction is applied for increasing the transmit power (i.e., following the equivalent isotropically radiated power (EIRP) requirement approved by Federal Communications Commission). In this work, the RF-WTP satisfying such two conditions is so-called the {\it green} RF-WTP. 
The characteristics of the green RF-WPT technique can be listed as follows
\begin{itemize}
\item The green RF-WPT technique plays a role as a bridge between green energy sources and energy-hunger devices.
\item The energy harvested at a receiver is foreseeable.
\item The amount of harvested energy belongs to transmit power, propagation loss, and wavelength.
\end{itemize}
Indeed, it is expected that the green RF resource is one of the most interesting candidates for future applications. 

\begin{figure}[t]
	\centering 
	{\includegraphics[width=0.45\textwidth]{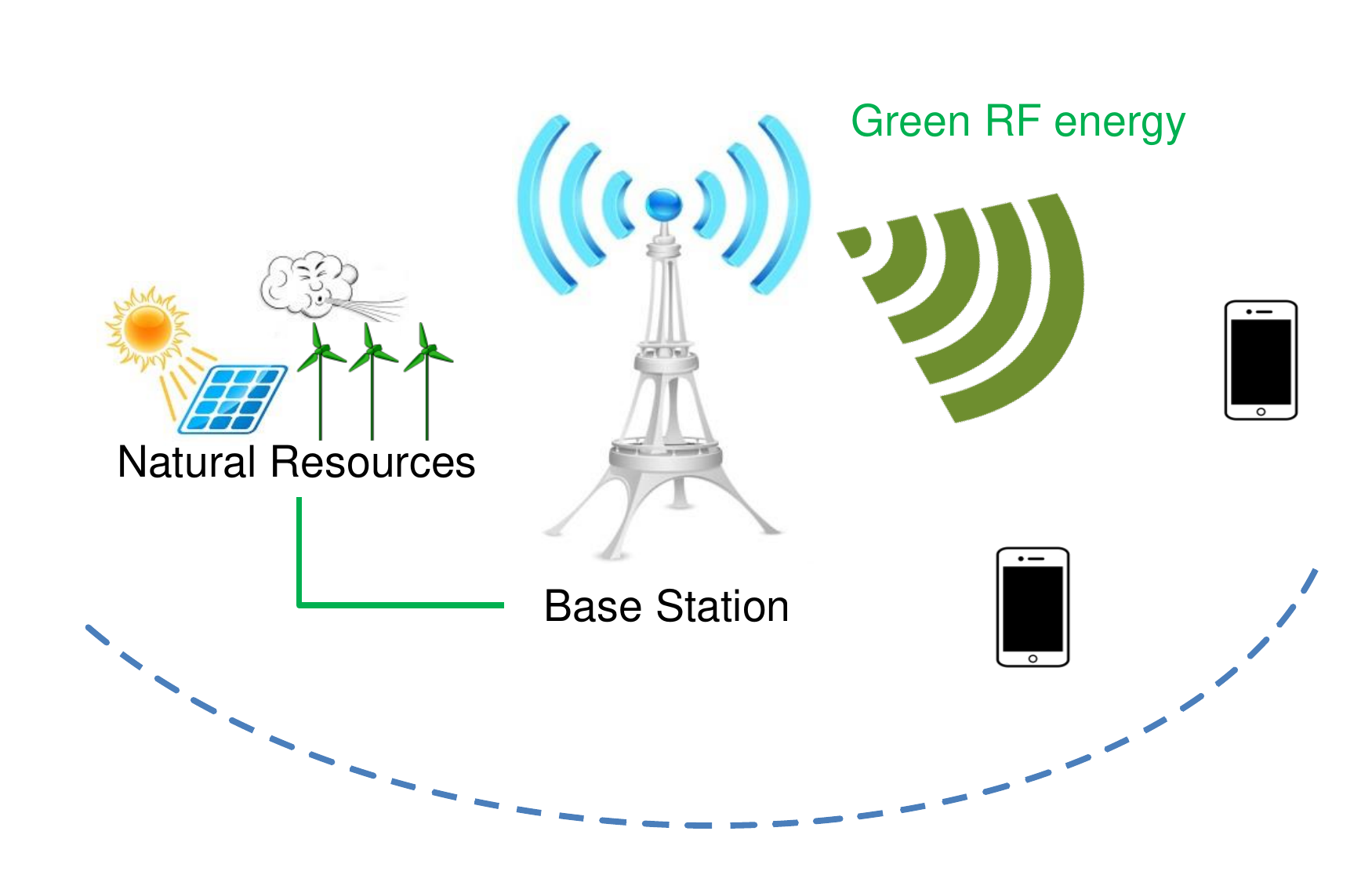}}
	\\ 
	{\caption{Green RF-WPT.}\label{fig:EHR}} 
\end{figure}

\section{A Vision of Future Green and EH Networks}

\subsection{A predicted model of future green networks with EH}
\begin{figure*}
	\centering 
	\includegraphics[width=0.7\textwidth]{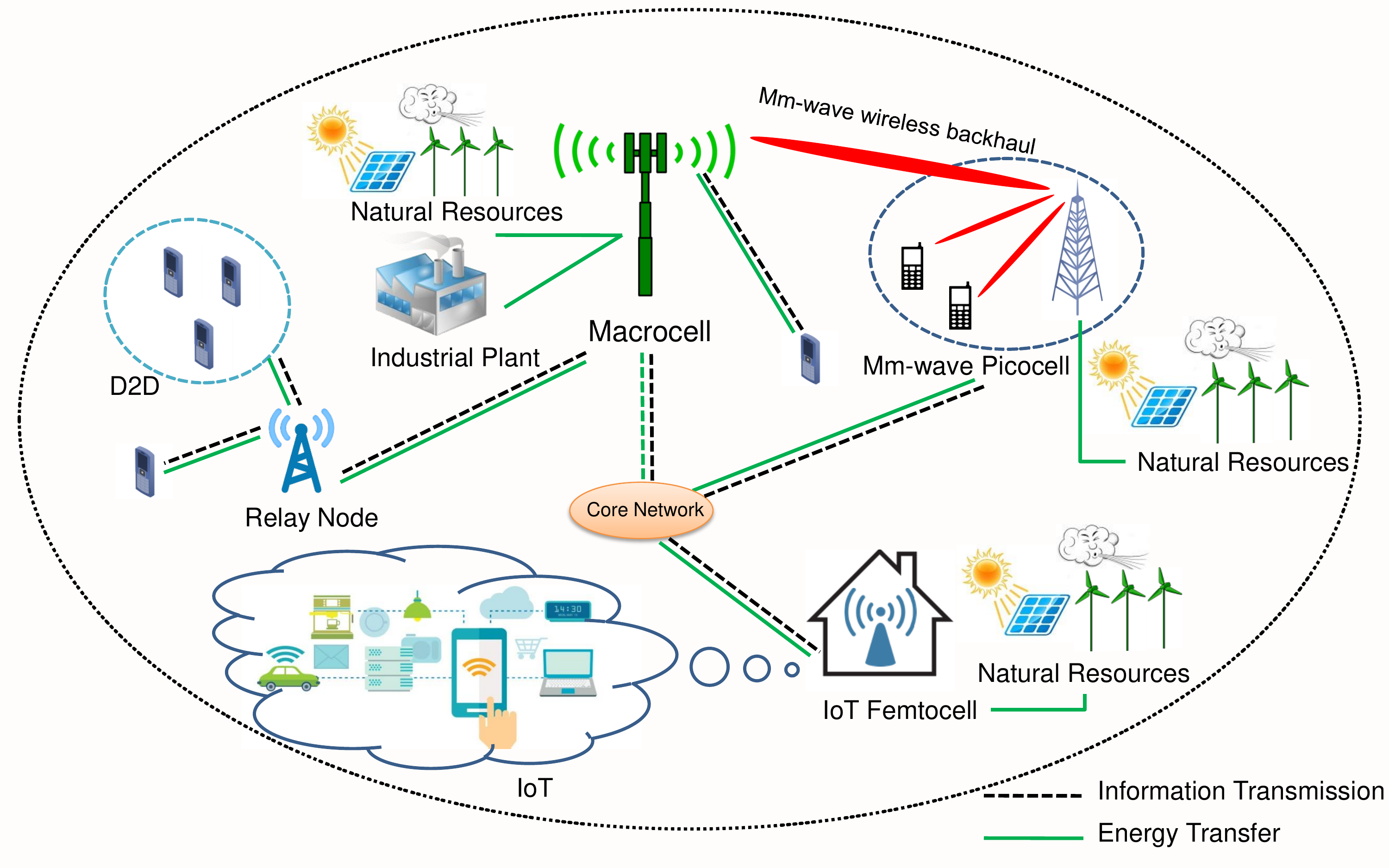}
	\\ 
	\parbox{0.75\textwidth}{\caption{A Future Green Network.}\label{fig:FGN}} 
\end{figure*}

 Future networks, e.g., 5G, are expected to support multimedia applications to achieve 1000-fold higher throughput, 1000-fold higher mobile data per unit area, and 10-fold longer lifetime of devices over the fourth generation (4G) networks [4], [5]. 
To adapt this progress, the design of new cellular networks tends to a new form embracing a large-scale deployment of small-cells.
Generally, the small-cells can be classified into distinct types including femtocell, microcell and picocell. 
Indeed, the multi-tier HetNet attains a promising gain in terms of spectral and energy efficiencies due to low power consumption and good ubiquitous connectivities [8].

On the other hand, nowadays,  the  development  of  wireless  networks  has  broken  the  limits  of 
power  consumption,  especially  in  cellular  networks.  Moreover,  the  energy 
cost  and  CO$_2$  emissions  have  been  promptly  growing  due  to  the  network 
operation.  Specifically,  this  has  inspired  researchers  with  a  challenging 
topic  so-called  future  green  wireless  networks.
As a promising solution, EH techniques exploit natural sources, and then contribute to reducing the overall footprint and extending the network lifetime. Nevertheless, the natural resources may not be always available to all devices. For instance, it is difficult for indoor devices to harvest solar energy. This yields another trend that BSs connected with outdoor harvesters which can harvest-and-store green energy when natural resources are available. Afterwards, BSs use such energy to wirelessly charge user devices using RF signals. 

Another approach to future green networks is the concept of green IoT [10]. In fact, IoT is an emerging trend that billions of identified low-power devices, e.g., sensor nodes, are connected to each other without the need for human interaction.
It can provide solutions to cut CO$_2$ emissions, reduce electromagnetic pollutions and improve energy efficiency. For instance, with tracking of motion sensors, the lights in rooms where there is no one inside would be turned off. Also, the green IoT technology can monitor energy usage in hi-tech buildings to reduce wasted energy. In fact, the green IoT is expected to enhance all technical, economical and environmental benefits. In particular, IoT network architectures mainly rely on the platforms of wireless sensor networks (WSNs), and cooperative networks to connect devices together [9]. In this concern, battery recharging for a large number of IoT devices is challenging indeed. 
Therefore, enabling the green IoT concept towards future networks requires advanced solutions of prolonging device's life-time, resource managements, and energy-efficient communication protocols.

Taking all the problems of interest into account, in the following, we further discuss several potential concepts towards green future networks as shown in Fig. 2. Specifically, challenges in implementing each concept are identified.

\subsection{Green radio communications: Main concepts and discussions}
\subsubsection{Full-duplex networks}
as mentioned, indoor devices which might not harvest green energy directly can be wirelessly powered by RF signals sent from BSs. Accordingly, this has inspired a combination of the full-duplex and simultaneous wireless information and power transfer (SWIPT) techniques. At the same time, the devices can receive energy in the downlink transmission while conveying information in the uplink connections to boost spectral efficiency.

Furthermore, we discuss two potential research issues of full-duplex SWIPT systems. First, the antennas at the full-duplex node are conventionally divided into transmit and receive sets. To improve the performance of SWIPT systems, an advanced form of the full-duplex technique that each antenna can simultaneously send information/energy and receive energy/information in the same frequency band is highly desirable. Indeed, this approach mainly depends on new hardware designs and new innovation of self-interference cancellation techniques. Second, full-duplex SWIPT small-cell base station can provide a promising approach regarding wireless backhauls in HetNets. In downlinks, the small-cell BS can receive information from macro-cell and transmit information/energy to users simultaneously. In uplinks, the small-cell BS can receive information/energy and send information to macro-cell at the same time. On this basis, the small-cell BSs do not require a separate frequency band of backhaul connections. Hence, resources and implementing costs are reduced.

\subsubsection{Millimeter-wave networks}
benefitting from mm-wave transmission, the overall electromagnetic field (EMF) exposure and power consumption per bit transmitted of networks are reduced due to a higher free-space attenuation at the mm-wave frequency, high directive antennas, and short distance links. 
Therefore, mm-wave communication has been considered
as a primary candidate for green cellular networks in future. 
This approach is expected to achieve multi-gigabit data rates
due to large spectrum resources at a ultra-high frequency band.
Specifically, EH devices can extract energy from incident RF
signals. Moreover, in mm-wave systems, many BSs are
densely deployed to ensure proper coverage to ultra-high frequency
networks. Thus, this can be attractive for the EH devices to potentially harvest sufficient energy.

In ultra-high frequency bands, the mm-wave signals mainly
suffer from the propagation loss, such as poor penetration
and diffraction. To make mm-wave networks more
favorable for SWIPT, beamforming techniques can be a promising
solution to increase the network coverage and system performance.
Moreover, in mm-wave networks, although the
small wavelength signals allow large antenna arrays 
to offer high beamforming gains, they require the alignment
between transmit and receive beams to reach the highest possible performance.
In these concerns, there is a non-trivial problem of the
beamwidth design. In practice, the length of beam-searching overhead
is directly proportional to the number of beamformer
candidates. The narrower beamwidth is, the more number of the beamformer candidates is, and then the longer
overhead is. This leads the time of data transmission to be
decreased. In constrast, the wider beamwidth means that it
is easier for transmit and receive beams to be aligned, and the
beam-searching process is sped up, however, the beamforming
gain is reduced. Therefore, future works should consider the
impact of beamwidth in various contexts to maximize the
SWIPT system performance.

\subsubsection{Wireless sensor networks}
over the last few years, the trend for WSNs is one of the most attractive
topics due to flexible installation
and convenient maintenance. Accordingly, many standards such as
WirelessHART, WIA-PA, and ISA100.11a have been proposed. 
Particularly, the IoT technology is mainly implemented on the WSN platform.  With an integration of IoT with sensors, sensor devices can be interconnected with the global Internet in order to provide solutions for future networks, such as reducing wasted energy in hi-tech buidings.

Specifically, replacing or charging the batteries in IoT WSNs may take time
and costs due to a large number of sensors, and this process
becomes dangerous for humans in hazardous environments. 
As a result, the EH from natural resources and RF signals for
WNSs have been considered as a promising
solution to prolong the sensor's lifetime. 

As the main distinction from conventional WSNs, EH-WSNs require new criteria in the fairness between information transfer and EH requirements.
In fact, the network can fail to adapt the EH
requirement while ensuring other system performances, such as throughput, delay or packet
loss.
As a result, leveraging between data transmission and EH is one of the critical concerns in designing EH-WSNs.
Therefore, efficient resource allocation schemes should take such a problem into account
to achieve high energy efficiency for EH-WSNs.

\subsubsection{Cooperative relay networks}
in recent, 
the cooperative relay network have been evaluated as one of the main core networks for the IoT technology where IoT nodes can communicate with each other and forward information and energy to the remote nodes. 
Up to now, many mature research works of cooperative communication have clearly shown
that the relay can be implemented not only to extend the
coverage range but also to improve the performance of wireless
communications. 
Further, the concept of EH/WPT relay networks
has been proposed and studied to enhance the lifetime of devices and overall performance of wireless networks. 
In cooperative EH/WPT relay networks, improving performance gain on the physical layer is one of the main research directions.
Hence, most previous efforts attempt to design novel schemes regarding relay operation policies, power allocation, precoder optimization and relay selection.

Considering existing challenges, the enhancement for both spectral and energy efficiencies in cooperative relay networks with the green RF-WTP is remarkable. In this concern, full-duplex or two-way relaying methods might be a promising solution.
Further, it is suggested that developed resource allocation schemes should consider the influence of incomplete channel state information (CSI) (e.g., the relay nodes have a partial user's CSI), and
the energy status at the relay nodes and users (e.g., the available energy, current power consumption, predicted energy harvested from natural resources or RF signals, etc.) on system performance.

\section{Future Research Issues}
In the prior section, several challenges of each concept have been presented. 
In future networks, since wireless communication systems are expected to be a mixture of various novel
system concepts to enhance both the spectral and energy efficiencies, some interesting combinations of the existing concepts are discussed as follows.

\subsection{When full-duplex communications meet mm-wave SWIPT networks?}

\begin{figure}[t]
	\centering 
	{\includegraphics[width=0.48\textwidth]{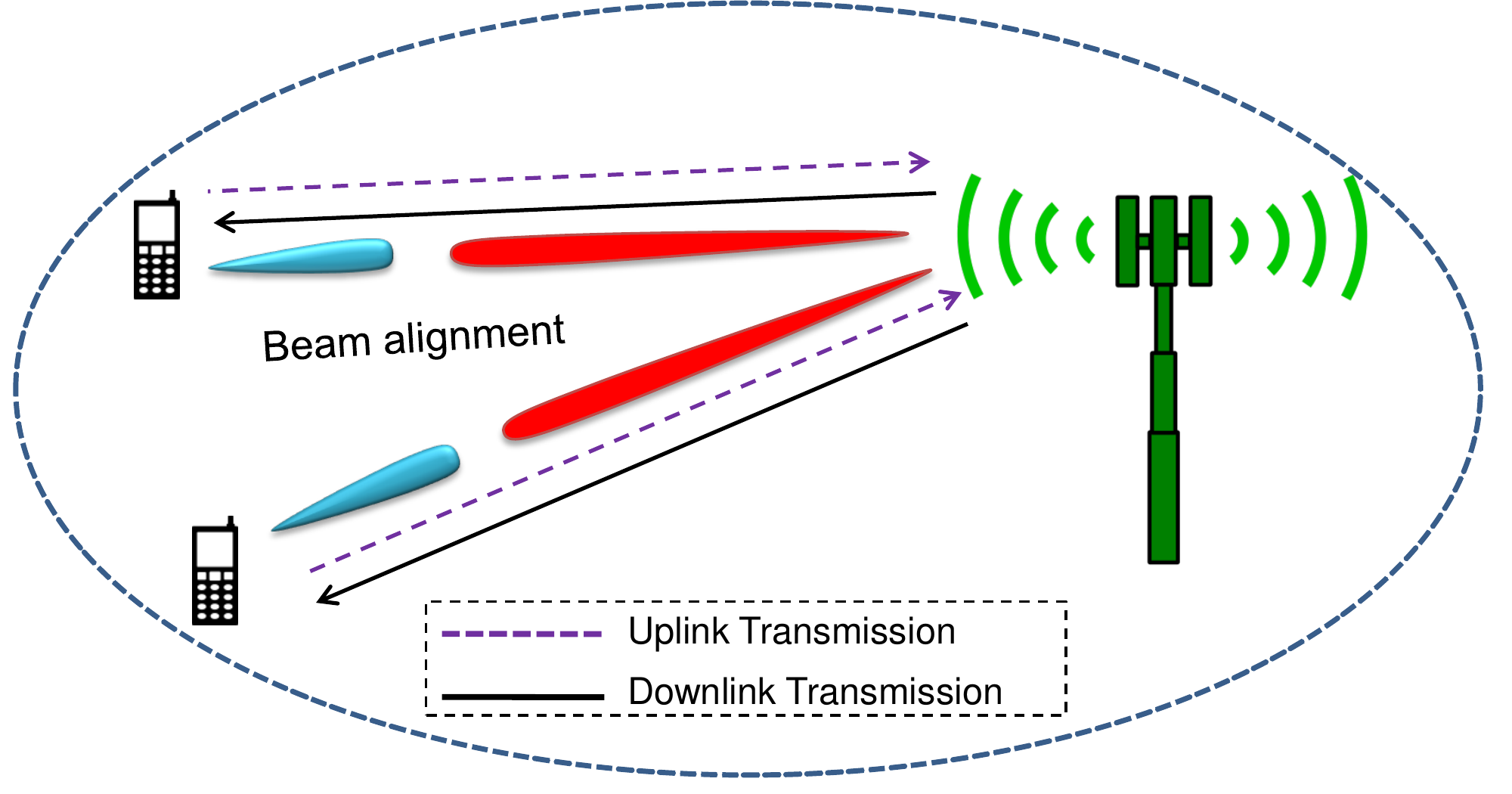}}
	\\ 
	{\caption{Mm-wave SWIPT networks with full-duplex communications.}\label{fig:FW1}} 
\end{figure}

A combination of mm-wave and full-duplex
SWIPT systems, as shown in Fig. \ref{fig:FW1}, seems to be interesting. Most of the recent
research on full-duplex SWIPT systems mainly address the
communications in conventional frequency bands. However,
in mm-wave frequency bands, there are two main challenges
need to be discussed. First, the practical implementation of
mm-wave full-duplex SWIPT should be investigated with a
bandwidth of several GHz. Second, in mm-wave networks,
the communication is inherently directional. Therefore, at both
the transmission and reception sides, the directional antenna
should be used. As a result, the node structure should be
re-considered according to the characteristics of mm-wave
signals. To reduce the cost of antennas (i.e. about parabolic
antennas), one of the efficient solutions is to employ transmit
and receive beams to limit self-interference.
Therefore, the future objectives should address:
\begin{itemize}
\item studying how the beamwidth affects the beam alignment, beamforming gain and beam-searching process,

\item investigating the performance trade-off between self-interference and data transmission time,

\item and then properly allocating available resources to optimize system performance.
\end{itemize}

\subsection{What are potential scenarios for SWIPT and EH HetNets with the full-duplex technique?}
\begin{figure}[t]
	\centering 
	{\includegraphics[width=0.48\textwidth]{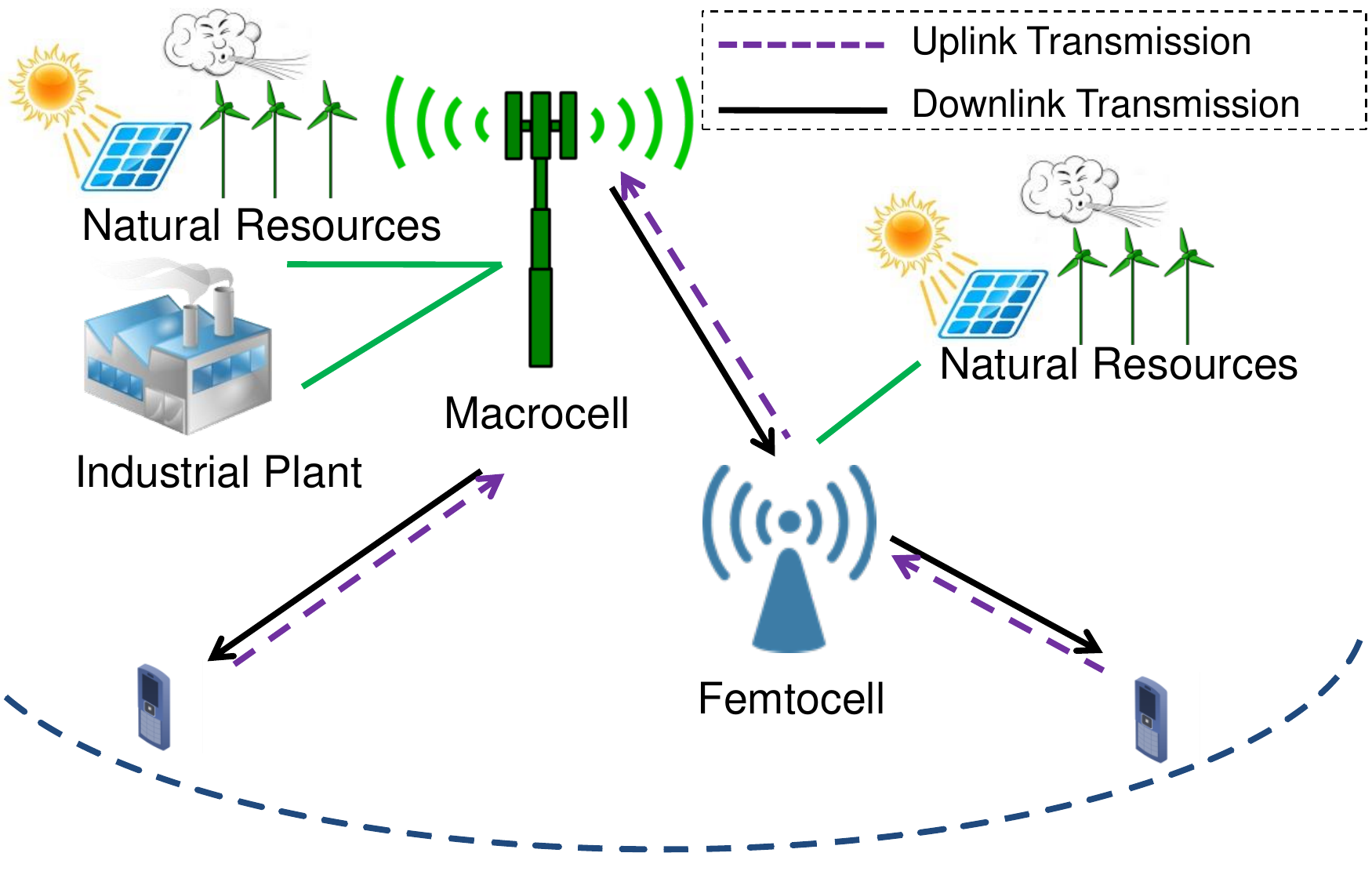}}
	\\ 
	{\caption{SWIPT and EH HetNets with the full-duplex technique.}\label{fig:FW2}} 
\end{figure}

Full-duplex EH-SWIPT HetNets may bring a bright, however, challenging approach. Given this concern, in Fig. \ref{fig:FW2},
a macro-cell BS harvests energy from natural sources and then communicates with a small-cell BS. On the other hand,
whereas the small-cell BS receives information from the macro-cell, it transmits information/energy to the users at the same time.
In another case, the small-cell BS transmits information to the macro-cell while it receives information from the user simultaneously.
Given these scenarios, the sytem benefits from an enhanced spectral efficiency, however, there appear a lot of interference sources, e.g., self-, inter-cell and intra-cell interferences due to full-duplex communications.
Specifically, dealing with downlink-to-uplink interference is a big challenge since the downlink power dominates the uplink one in general.
Therefore, using optimization frameworks, future works need to focus on: 
\begin{itemize}
\item designing new self-interference cancellation techniques,

\item managing the downlink-to-uplink interference,

\item allocating resources to optimize the system performance in terms of the information and power transfer.
\end{itemize}

\subsection{What are the main concerns of wirelessly powering Internet of Things networks?}
\begin{figure}[t]
	\centering 
	{\includegraphics[width=0.48\textwidth]{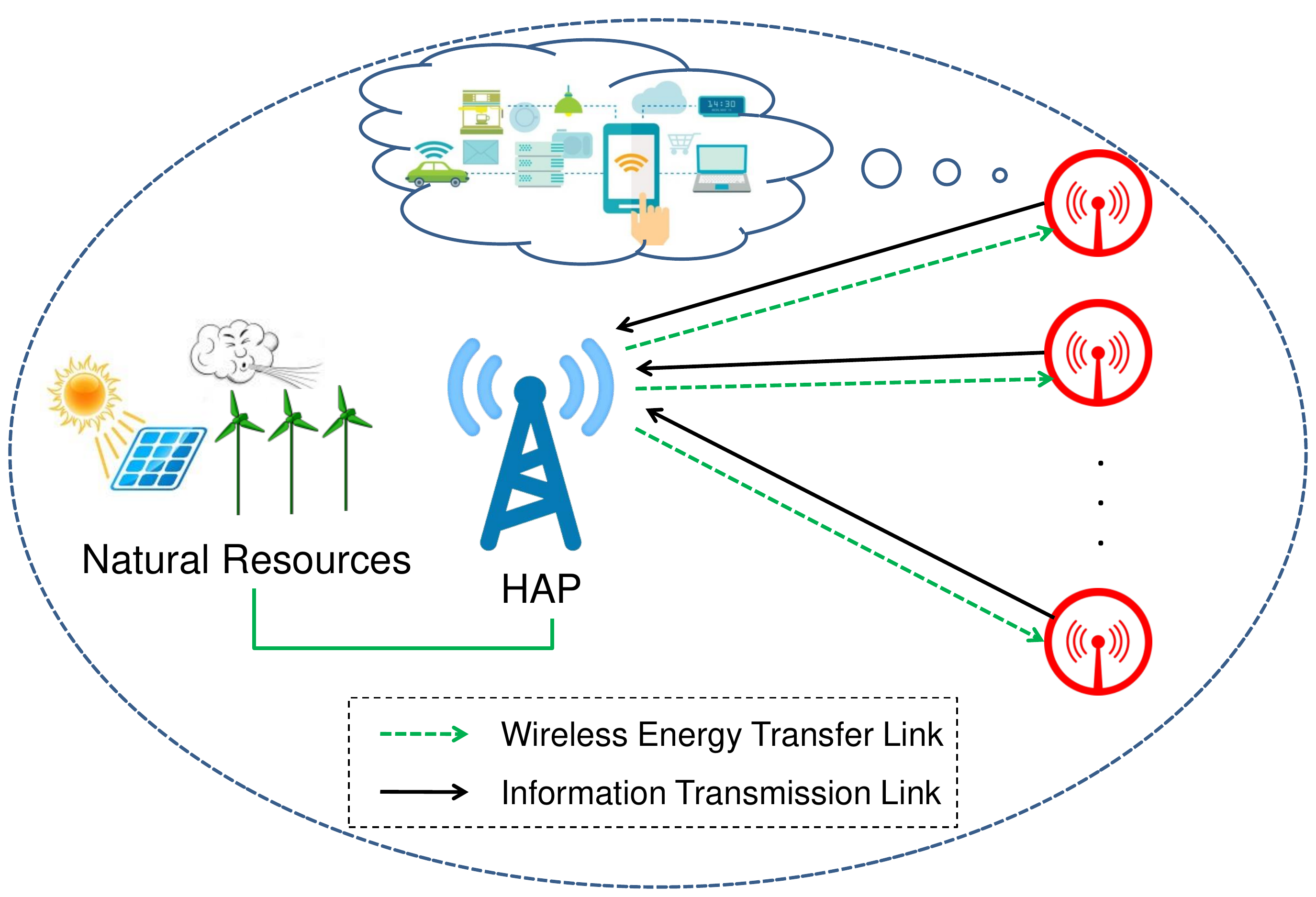}}
	\\ 
	{\caption{RF-WPT Internet of Things networks.}\label{fig:FW3}} 
\end{figure}

Due to large-scale deployments of IoT networks, replacing or recharging the device's battery is one of the main challenges.
Specifically, a large number of IoT sensors are implemented in indoor locations where natural resources might be not available to harvest directly. In this context, the green RF-WPT is a promising candidate for prolonging the lifetime of the IoT low-power devices.
Thus, this implies a scenario as illustrated in Fig. \ref{fig:FW3} where a sink node is responsible for harvesting energy from natural resources and then wirelessly transferring power to devices in a IoT wireless sensor network. In addition, the devices can communicate with each other. 
Given the model, some important concerns need to be addressed as follows.
\begin{itemize}
\item scheduling the power transfer to energy-hungry users according to the harvested energy at the sink node,

\item exploiting interference from ambient environments to improve the EH performance,

\item maximizing the information performance while satisfying EH requirements.
\end{itemize}

\section{Concluding Remarks}
In this work, we have presented a review of promising trends towards future green networks. Based on the platform of EH techniques, several potential concepts such as HetNet, mm-wave and IoT networks, have been presented. In particular, we discuss a promising architecture, so-called green RF-WPT. In fact, the latter plays a crucial role as a bridge between natural energy resources and smart energy-hungry devices. 
Accordingly, we have shown a vision of future green networks in which smart devices can be recharged by green resources even when they cannot harvest energy directly.
Furthermore, to facilitate the regreening process while adopting intensive system performance required in future networks, the combinations of techniques, such as SWIPT, mm-wave, full-duplex, can bring outstanding outcomes. Given this concern, we have identified some challenges in mixing the potential concepts, and discussed how they can work together.
Indeed, it is expected that the green RF-WPT-based approaches can be one of the potential solutions for regreening the future ICT world.

\begin{IEEEbiographynophoto}{Ha-Vu Tran}
received a bachelor degree in Electronic
and Telecommunication Engineering from
Hue University of Sciences, Vietnam in 2012. In
2015, he completed master degree in Electronics
and Radio Engineering from Kyung Hee University,
South Korea. Currently, he is pursuing his Ph.D.
degree at \'{E}cole de Technologie Sup\'{e}rieure (ETS),
University of Qu\'{e}bec, Canada.
\end{IEEEbiographynophoto}

\begin{IEEEbiographynophoto}{Georges Kaddoum} (M'11)
is an associate Professor
of electrical engineering with the \'{E}cole de Technologie Sup\'{e}rieure (ETS), University of Qu\'{e}bec, Montr\'{e}al, QC, Canada. He received his B.Sc. degree from the \'{E}cole Nationale Sup\'{e}rieure de Techniques Avanc\'{e}es (ENSTA Bretagne), Brest, France, and the M.S. degree from the Universit\'{e}
de Bretagne Occidentale and Telecom Bretagne (ENSTB),
Brest, in 2005 and the Ph.D. degree (with
honors) from the National
Institute of Applied Sciences (INSA), University of Toulouse, Toulouse,
France, in 2009.
\end{IEEEbiographynophoto}
\end{document}